\documentclass[aps,letterpaper,twocolumn,preprintnumbers,floatfix,superscriptaddress,nofootinbib]{revtex4-1}
\pdfoutput=1
\usepackage{graphicx}  
\usepackage{dcolumn}  
\usepackage{bm}        
\usepackage{amssymb}
\usepackage{amsmath}
\usepackage{slashed}  
\usepackage[colorlinks=true,linkcolor=red,anchorcolor=black,citecolor=blue,filecolor=cyan,menucolor=red,runcolor=filecolor,urlcolor=blue,bookmarks=true,bookmarksnumbered=true]{hyperref}
\usepackage{bbold}

\usepackage[usenames,dvipsnames,svgnames]{xcolor}
\usepackage[english]{babel}
\usepackage[applemac]{inputenc}
\usepackage{color}
\definecolor{rossos}{cmyk}{0,1,1,0.55}
\definecolor{bluscuro}{rgb}{0.15, 0.2, .85}
\definecolor{bluchiaro}{cmyk}{1,.3,0.,0.1}
\definecolor{grey}{rgb}{0.6,0.6,0.6}
\definecolor{fuchsia}{rgb}{1,0,1}

\newcommand{\beq}{\begin{equation}}
\newcommand{\eeq}{\end{equation}}
\newcommand{\bea}{\begin{eqnarray}}
\newcommand{\eea}{\end{eqnarray}}
\newcommand{\eq}[1]{eq.\!~(\ref{#1})}
\newcommand{\eqs}[1]{eqs.\!~(\ref{#1})}

\newcommand{\ie}{$\textnormal{i.e.}$}

\newcommand{\GeV}{\,\mathrm{GeV}}
\newcommand{\TeV}{\,\mathrm{TeV}}

\newcommand{\vev}[1]{\langle {#1} \rangle}

\newcommand{\LUV}{\Lambda_{\slash \hspace{-5pt} Z_2}}

\begin{document}

\widetext
\leftline{CERN-TH-2017-192}

\title{The Brother Higgs\vspace{2mm}\\
\tiny The missing sibling of the Higgs}
\author{Javi Serra}
\affiliation{CERN, Theory Department, CH-1211 Geneve 23, Switzerland\vspace{0.5mm}}
\author{Riccardo Torre\vspace{1.5mm}}
\affiliation{INFN, Sezione di Genova, Via Dodecaneso 33, I-16146 Genova, Italy}

\begin{abstract}
We present a version of the twin Higgs mechanism with minimal symmetry structure and particle content.
The model is built upon a composite Higgs theory with global $SO(6)/SO(5)$ symmetry breaking.
The leading contribution to the Higgs potential, from the top sector, is solely cancelled via the introduction of a Standard Model neutral top partner.
We show that the inherent $Z_2$ breaking of this construction is under control and of the right size to achieve electroweak symmetry breaking, with a fine-tuning at the level of 5-10\%, and compatibly with the observed Higgs mass.
We briefly discuss the particular phenomenological features of this scenario.
\end{abstract}


\maketitle


\section{Introduction} \label{sec:intro}

The LHC has made the electroweak (EW) hierarchy problem a concrete issue.
Supersymmetry (SUSY) or Compositeness, the long-term solutions to this naturalness problem, require colored particles within LHC reach and therefore seem increasingly disfavored by data.
However, clever modifications of the standard scenarios can still provide rather natural solutions.
This happens, for instance, within the neutral-naturalness paradigm: the little fine-tuning problem of SUSY or composite theories is cured via the introduction of light Standard Model (SM) neutral states, pushing the scale of the SUSY or composite partners beyond LHC reach.

Twin Higgs (TH) models are an explicit realization of this paradigm.
The Higgs arises as a Nambu-Goldstone boson (NGB) from the spontaneous breaking of a global symmetry $\mathcal{G}$ to $\mathcal{H}$, at a scale $f \approx \TeV$. 
As a consequence, Higgs non-derivative interactions, and in particular its potential, are protected by symmetry.
The couplings of the Higgs to the SM fermions and gauge bosons break explicitly such a Higgs shift-symmetry, inducing a sensitivity of the radiatively generated Higgs potential to the cutoff scale $m_* \sim g_* f \gtrsim \TeV$, \ie~the mass of the SUSY or composite partners.
However, in TH models this sensitivity can be weakened via the introduction of neutral copies, called twins, of the SM fields \cite{Chacko:2005pe,Barbieri:2005ri,Chacko:2005vw,Chacko:2005un,Chang:2006ra,Batra:2008jy,Craig:2013aa,Craig:2014aea,Geller:2014kta,Burdman:2014zta,Craig:2014roa,Craig:2015pha,Barbieri:2015aa,Low:2015aa,Curtin:2015bka,Csaki:2015gfd,Craig:2016kue,Harnik:2016koz,Barbieri:2016zxn,Greco:2016zaz,Chacko:2016hvu,Craig:2016lyx,Katz:2016wtw,Contino:2017moj}. 
This allows to realize a heavier spectrum of SM-charged (in particular colored) partners without worsening fine-tuning.
Indeed, a discrete $Z_2$ symmetry that relates the Higgs as well as the SM matter and gauge particles to their corresponding twins,
ensures that the leading order (LO) potential, proportional to $m_*^2$, is $\mathcal{G}$-invariant, therefore Higgs independent. 
In this letter, we wish to identify the essential features of this mechanism, by keeping only its bare minimum ingredients, at least from a low-energy perspective.

The key player in the radiative generation of the Higgs potential is the top quark, while the EW gauge and scalar contributions are not 
a severe concern:
given the energies accessible at the LHC, the latter do not lead to a significant fine-tuning problem.
Therefore the indispensable element in TH constructions is the SM-neutral partner of the top quark (charged under a copy of QCD), together with a parity that guarantees a vanishing LO contribution to the Higgs potential from the top sector.
Given the top Yukawa, \mbox{$y_t \bar t_R H q_L + \mathrm{h.c.} = y_t \bar t_R (h^0 t_L - h^+ b_L ) + \mathrm{h.c.}$}, then all that is needed is that the copy of the top, $\tilde t_{L,R}$, couples to a complex scalar $\tilde h^0$, $\tilde y_t \bar{\tilde t}_R \tilde h^0 \tilde t_L + \mathrm{h.c.}$, with strength $\tilde y_t = y_t$. 
This would suggest that the global symmetry that relates $H$ and $\tilde h^0$ is $\mathcal{G} = SU(3) \times U(1)$ \cite{Poland:2008ev,Cai:2008au}, however a NGB Higgs cannot arise from its spontaneous breaking while preserving a custodial $SU(2)_L \times SU(2)_R \times P_{LR} \supset \mathcal{H}$ symmetry \cite{Sikivie:1980hm,Agashe:2006at}. Since the smallest symmetry group $\mathcal{G}$ under which this Higgs sector transforms should have at least rank three, then the minimal custodially protected symmetry breaking pattern where the cancellation of the LO top potential can be realized is $SO(6)/SO(5)$. In this paper we focus on this case.%

Importantly, by construction a $Z_2$ symmetry that relates $t_{L,R} \leftrightarrow \tilde t_{L,R}$ and $h^0 \leftrightarrow \tilde h^0$ can only enforce $\tilde y_t = y_t$ up to explicit breaking effects mediated by the bottom quark or the gauge bosons. Despite these breakings seem to prevent a proper implementation of the parity, they can be regarded as NLO effects, against which TH models do not provide a protection of the Higgs potential anyway.
Moreover, even if one might conclude that the absence of a bona-fide discrete symmetry precludes $\tilde y_t \approx y_t$ to start with, one can in fact think of completions where an exact $Z_2$ symmetry gets broken at a high UV scale $\LUV >  m_*$, delivering our model at lower energies
(we provide examples of such completions in Appendix~\ref{app:twin}).
We carefully study under which conditions running effects down to the scale $m_*$, which tend to split the Yukawa of the top and the Yukawa of what we call the \emph{brother} top, $\tilde t$, can be kept under control.%
\footnote{This is somewhat reminiscent of models of gauge coupling unification (or Yukawa coupling unification): assuming the existence of the brother top coupled to an $SO(6)$-symmetric Higgs sector, the relation $\tilde y_t \approx y_t$ at the scale $m_*$ is inferred from the fact that the Higgs potential is small, suggesting a common origin of both couplings in the UV, even if an exact symmetry is absent in the IR. 
By running the Yukawa couplings to high energies, one can estimate the scale at which such a symmetry should be recovered.}
Provided this is the case, the tuning in our scenario is approximately given by twice $\xi = v^2/f^2$ with $v \approx 246 \GeV$, as in standard TH models.

The paper is organized as follows. 
In Section~\ref{sec:higgs} the structure and symmetries of the sector giving rise to the Higgs are presented. 
Sections~\ref{sec:gauge} and \ref{sec:fermion} are devoted to the gauge and fermionic sectors respectively, as well as to their contributions to the Higgs potential. 
The breaking of the EW symmetry and the generation of the Higgs mass are discussed in detail in Section~\ref{sec:ewsb}. In Section~\ref{sec:pheno} we outline the phenomenological implications of our model, paying special attention to the differences with respect to the usual TH phenomenology.
Conclusions are drawn in Section~\ref{sec:conclusion}. 


\section{The Strong Sector} \label{sec:higgs}

We base our discussion on a composite Higgs model, with mass gap $m_*$ and typical coupling between the composite states $g_*$.
The Higgs doublet emerges as a NGB, along with an EW singlet $\eta$, from a strong sector's global $SO(6)$ symmetry spontaneously broken to $SO(5)$ \cite{Gripaios:2009pe,Serra:2015xfa}. 
We chose to parametrize the $SO(6)/SO(5)$ coset with the non-linear vector
\bea
&\Phi = U(\pi) \Phi_0 = \begin{pmatrix}
\pi_1 & \pi_2 & \pi_3 & \pi_4 & \pi_5 & \sigma
\end{pmatrix}^T \,, \\
&\pi_4 \equiv h \, , \, \pi_5 \equiv \eta \,, \, \sigma = \sqrt{1-\pi_{\hat a}^2} \,, \nonumber 
\label{Phi}
\eea
where $\hat a = 1, \dots, 5$.
The NGBs transform as a $\mathbf{5}$ of $SO(5)$, while $\Phi \sim \mathbf{6}$ of $SO(6)$.
The matrix of Goldstones $U(\pi)$, as well as the fifteen generators $T^A$ of $SO(6)$, can be found in Appendix~\ref{app:so6}.
The scalar $\eta$, along with the putative radial excitation of the symmetry breaking scale, identified with $\sigma$, conform what we call the \emph{brother} Higgs (referred to as $\tilde h^0$ in the introduction).
The Higgs is given by $H = \begin{pmatrix} h^+ & h^0 \end{pmatrix}^T = {1 \over \sqrt{2}} \begin{pmatrix} \pi_2 + i \pi_1 & h - i \pi_3 \end{pmatrix}^T $.

The kinetic term for the NGBs, in the absence of any gauging of the $SO(6)$ symmetries, is simply given by
\beq
\frac{f^2}{2} |\partial_\mu \Phi|^2 = \frac{f^2}{2} (\partial_\mu \pi_{\hat a})^2 + \frac{f^2}{2} \frac{(\pi_{\hat a} \partial_\mu \pi_{\hat a})^2}{1-\pi_{\hat b}^2} \,,
\label{kinterm}
\eeq
where $f \sim m_*/g_*$ is the $\sigma$-model decay constant.

Among the \emph{internal} parities of $SO(6)$, we are particularly interested in 
\beq
\mathcal{P} = \begin{pmatrix}
\mathbb{1}_2 &  &  \\
  &  & \mathbb{1}_2 \\
  & \mathbb{1}_2 & 
\end{pmatrix} \,,
\label{parity}
\eeq
which exchanges $\begin{pmatrix} \pi_3 & h \end{pmatrix}$ with $\begin{pmatrix} \eta & \sigma \end{pmatrix}$. 
Notice that $\pi_3$ is the NGB eventually eaten by the $Z$ boson, after gauging the EW subgroup, while the $W^\pm$ eat $\pi_{1,2}$.

The strong sector is also assumed to preserve global $SU(3)_C \times U(1)_X$ and $SU(3)_{\widetilde C} \times U(1)_{\widetilde X}$ symmetries, as well as an \emph{external} parity $\mathcal{Z}_2$ exchanging $C, X \leftrightarrow \widetilde C, \widetilde X$.
The combined action of $\mathcal{P} \times \mathcal{Z}_2$ on the strong sector realizes the $Z_2$ symmetry required to implement the TH mechanism on the top sector.

Beyond the conserved currents associated to its global symmetries, $\mathcal{J}_\mu^A \sim {\bf 15}$ of the $SO(6)$, $\mathcal{J}_\mu^{C, X}$ and $\mathcal{J}_\mu^{\widetilde C, \widetilde X}$, the strong sector also contains fermionic operators with non-trivial quantum numbers under $SO(6) \times SU(3)_C \times SU(3)_{\widetilde C} \times U(1)_X \times U(1)_{\widetilde X}$, in particular $\Psi_s \sim ({\bf 1},{\bf 3},{\bf 1})_{2/3,0}$ and $\widetilde \Psi_s \sim ({\bf 1},{\bf 1},{\bf 3})_{0,2/3}$, as well as $\Psi_v \sim ({\bf 6},{\bf 3},{\bf 1})_{2/3,0}$ and $\widetilde \Psi_v \sim ({\bf 6},{\bf 1},{\bf 3})_{0,2/3}$ with $\Psi_{s,v} \leftrightarrow \widetilde \Psi_{s,v}$ under $\mathcal{Z}_2$.
The $SO(6)$ fermionic multiplets decompose under the unbroken $SO(5)$ as ${\bf 6} = {\bf 5} \oplus {\bf 1}$, such that $\Psi_v = \begin{pmatrix} \Psi_{\bf 5} & \Psi_{\bf 1} \end{pmatrix}^T$ and likewise for $\widetilde \Psi_v$. 
The $SO(6)$ vector current decomposes as ${\bf 15} = {\bf 5} \oplus {\bf 10}$.


\section{The Gauge Sector} \label{sec:gauge}

A subgroup of the global symmetries of the strong sector is gauged by elementary vector fields.
Within $SO(6 ) \times U(1)_X$, the subgroup $SU(2)_L \times U(1)_Y$, with $Y = T_R^3 + X$, is identified with the EW group.
Likewise, $SU(3)_C$ is identified with the QCD group.
Beyond the SM gauge content, we can introduce two extra sets of elementary vectors that gauge $SU(3)_{\widetilde C}$ and $U(1)_{\widetilde Q}$ with $\widetilde Q = T_\eta/\sqrt{2} + \widetilde X$. 
We call these new gauge fields the brother gauge bosons.

Weak gauging implies linear couplings between the elementary gauge fields and the corresponding strong sector currents, \ie~$A_\mu \mathcal{J}^\mu$, that are reproduced by the covariant derivatives $\partial_\mu \to D_\mu = \partial_\mu - i A_\mu^A T^A$, where
\bea
A_\mu^A T^A &=& W_\mu^\alpha T_L^\alpha + B_\mu Y + G_\mu^a T_C^a \nonumber \\
&& + \, \widetilde Z_\mu \widetilde Q + \widetilde G_\mu^a T_{\widetilde C}^a 
\label{gaugeembedding}
\eea
identifies the embedding of the elementary gauge bosons ($\alpha = 1,2,3$, $a = 1, \dots, 8$).

The presence of kinetic terms independent of the strong dynamics is what characterizes the elementary gauge bosons,
\beq
-\frac{1}{4g_2^2} W_{\mu\nu}^\alpha W^{\mu\nu}_\alpha  -\frac{1}{4g_1^2} B_{\mu\nu} B^{\mu\nu} -\frac{1}{4 \tilde g^2} \widetilde Z_{\mu\nu} \widetilde Z^{\mu\nu} \,,
\eeq
and likewise for $G$ and $\widetilde G$ with gauge couplings $g_3$ and $\tilde g_3$, respectively.
Consistently with the $\mathcal{Z}_2$ symmetry of the strong sector, a discrete $Z_2$ symmetry $G^a_\mu \leftrightarrow \widetilde G^a_\mu$ can be imposed on the elementary sector, which enforces $g_3 \approx \tilde g_3$.
However, we note that in our model there is no limit in which a similar parity can be imposed between (any of) the EW gauge bosons and the $\widetilde Z$.
For instance, the internal parity $\mathcal P$ exchanges $T_A^3 \equiv (T_L^3 - T_R^3)/\sqrt{2}$ with $T_\eta$, but $T_L^\alpha$ and $T_R^3$ are gauged with two different gauge couplings, $g_2$ and $g_1$ respectively, with no obvious connection to $\tilde g$, at least from an IR perspective.
In fact, due to the absence, by construction, of an exact $Z_2$ symmetry in the EW gauge sector, other combinations $\widetilde Q = \alpha T_\eta + \beta \widetilde X$ could be gauged as well, or $\widetilde Q$ could even not be gauged at all.
In Appendix~\ref{app:twin} we expand on this discussion, providing plausible relations between the gauge couplings from the requirement that a complete $Z_2$ symmetry in the gauge sector is recovered at a high scale $\LUV > m_*$, as well as discussing the consequences of not gauging $\widetilde Q$, in which case $\eta$ remains uneaten.
Regardless of these considerations, we note that provided $\tilde g$ is not strong at $m_*$, such that the associated Higgs potential is under control (see \eq{gaugepot} below), the fact that $\tilde g$ is unrelated to the EW couplings does not pose any serious issue.

According to the gauging in \eq{gaugeembedding}, the $W^\pm$, $Z$ and $\widetilde Z$ become massive, with the required extra degrees of freedom, $\pi_\alpha$ and $\eta$, provided by the strong sector, and the required interactions following from \eq{kinterm}.
Canonically normalizing the gauge bosons and moving to the unitary gauge, where $\pi_\alpha = \eta = 0$, the complete kinetic term reads
\bea
\frac{f^2}{2} |D_\mu \Phi|^2 &=& \frac{f^2}{2} \frac{(\partial_\mu h)^2}{1-h^2} \nonumber \\
&& + \frac{g^2 f^2}{4} h^2 \left( W_\mu^+ W^{\mu -} + \frac{1}{2 c_{\theta}} Z_\mu Z^\mu \right) \nonumber \\
&& + \frac{\tilde g^2 f^2}{8} (1-h^2) \widetilde Z_\mu \widetilde Z^\mu \,.
\eea
Obviously, EW symmetry breaking (EWSB) must take place, \ie~$\vev{h}^2 = v^2/f^2 = \xi$, for the SM gauge bosons to become massive.\\

The partial gauging of the strong sector's $SO(6)$ global symmetry explicitly breaks the shift-symmetries associated with the Higgs, giving rise to a potential for the then pseudo-NGB $h$.
The leading order (LO) contribution, derived from symmetry considerations only, reads
\begin{align}
& V_{g^2}^{\textrm{UV}} = c_g \sum_A g_A^2 \Phi^T T^A T^A \Phi \nonumber \\
& \qquad = c_g \frac{1}{4} \left[ (g_1^2 + 3g_2^2) h^2 + \tilde g^2 (1-h^2) \right] \,,
\label{gaugepot}
\end{align}
where the coefficient $c_g$ is exactly the same for all the terms in the sum, owing to the symmetries of the strong sector.
Its NDA estimate is $c_g \sim 3 m_\rho^2 f^2/32\pi^2$, where $m_\rho \lesssim m_*$ can be interpreted as the mass of a composite vector resonance (with the quantum numbers of the strong sector current $\mathcal{J}_\mu^A$) regulating the size of the potential. 
Equation~(\ref{gaugepot}) is the only purely gauge contribution to the Higgs potential relevant for our discussion.
Note that the EW piece is partially cancelled by that of the brother $Z$.
The remaining $Z_{2}$-breaking terms will be important in the following, since they provide a Higgs mass term that is needed to achieve a phenomenologically viable EW minimum, as discussed in Section~\ref{sec:ewsb}.


\section{The Fermionic Sector} \label{sec:fermion}

The elementary sector includes fermionic fields with the quantum numbers of the SM fermions.
These interact with the strong sector via partial compositeness \cite{1991NuPhB.365..259K}, \ie~linear couplings with composite fermionic operators $\bar \psi \Psi$, at least for what regards the third generation quarks.
In particular, $q_L$ couples to $\Psi_v$ with $y_L$ strength, while $t_R$ couples to $\Psi_s$ with coupling $y_R$.
Note that when $y_R \to g_*$, limit we are interested in, $t_R$ can be directly identified with the right-handed component of $\Psi_s$, taken in this case as a (chiral) massless composite fermion. 
The low-energy interactions of the left-handed top (and bottom) are parametrized by the embedding
\beq
Q_L = v_{b} b_L + v_{t} t_L = \frac{1}{\sqrt{2}} 
\begin{pmatrix}
i \, b_L & b_L & i \, t_L & -t_L & 0 & 0
\end{pmatrix}^T \,.
\label{qembed}
\eeq
Two extra elementary fermions, the brother left- and right-handed tops, $\tilde t_L$ and $\tilde t_R$, are introduced.
They are singlets under the SM elementary symmetries, but carry charges under the extra gauged symmetries, specifically $\tilde t_L \sim {\bf 3}_{7/6}$ and $\tilde t_R \sim {\bf 3}_{2/3}$ of $SU(3)_{\widetilde C} \times U(1)_{\widetilde Q}$.
They couple to the $\mathcal{Z}_2$ counterparts of the composite fermions mixing with $q_L$ and $t_R$, \ie~$\widetilde \Psi_v$ and $\widetilde \Psi_s$, with strengths $\tilde y_L$ and $\tilde y_R$, respectively. 
The interactions of $\tilde t_L$ are then parametrized by the embedding
\beq
\widetilde Q_L = v_{\tilde t} \tilde t_L = \frac{1}{\sqrt{2}} 
\begin{pmatrix}
0 & 0 & 0 & 0 & i \, \tilde t_L & - \tilde t_L
\end{pmatrix}^T \,.
\label{qtwinembed}
\eeq
Both the top and brother top acquire Yukawa couplings,
\begin{align}
& y_t f \bar Q_L \Phi t_R + \tilde y_t f \bar{\widetilde Q}_L \Phi \tilde t_R + \text{h.c.} \nonumber \\ 
& \qquad = -\frac{y_t}{\sqrt{2}} f \bar{t}_L h t_R -\frac{\tilde y_t}{\sqrt{2}} f \bar{\tilde t}_L \sqrt{1-h^2}  \tilde t_R + \mathrm{h.c.} \,,
\label{Yukcouplings}
\end{align}
where selection rules and NDA fix $y_t \sim y_L y_R /g_{*}$ while $\tilde y_t \sim \tilde y_L \tilde y_R/g_{*}$.\\

Given that the couplings of $q_L$ and $\tilde t_L$ explicit break the shift-symmetry of the Higgs, a potential is radiatively generated.
The LO contribution, from one-loop top or brother top diagrams, can be derived based on their spurionic quantum numbers
\begin{align}
& V_{y^2}^{\textrm{UV}} = c_y y_L^2 \sum_{\psi=t,b} \Phi^T v_\psi v_\psi^\dagger \Phi + c_y \tilde y_L^2 \Phi^T v_{\tilde t} v_{\tilde t}^\dagger \Phi \nonumber \\
& \qquad = c_y \frac{1}{2} \left[ y_L^2 h^2 + \tilde y_L^2 (1-h^2) \right] \,,
\label{LOtoppot}
\end{align}
where the same coefficient $c_y$ is present for all the terms in the potential, owing to the symmetries of the strong sector: in particular the parity $\mathcal{P} \times \mathcal{Z}_2$, which effectively acts as $t_{L,R} \leftrightarrow \tilde t_{L,R}$. 
This coefficient is estimated as $c_y \sim 6 m_\Psi^2 f^2/32\pi^2$, where $m_\Psi \lesssim m_*$ can be interpreted as the mass of a composite fermionic resonance at which this term is saturated.
Crucial in our construction is the fact that when $y_L \approx \tilde y_L$ at the scale $m_\Psi$, the Higgs dependence of the potential in \eq{LOtoppot} approximately cancels. 
Note that the $b_L$ only plays the role of preserving the $SU(2)_L \times U(1)_Y$ invariance of the potential.
As in standard TH models, the twin bottom does not play any role in the aforementioned cancellation.
This is why there is no need to specify the gauge quantum numbers of $\tilde b_{L,R}$ or their couplings to the strong sector (as long as these are small), nor that any $Z_2$ symmetry is respected in the bottom sector.
Gauge anomalies associated with $SU(3)_{\widetilde C} \times U(1)_{\widetilde Q}$ should vanish. There are several ways to achieve this, even if no $Z_2$ partners for the leptons are present. For instance one can take $\tilde b_L \sim {\bf 3}_{-7/6}$ and $\tilde b_R \sim {\bf 3}_{-2/3}$ (from e.g.~$T_\eta(\tilde b_L) = -1/\sqrt{2}$, $T_\eta(\tilde b_R) = 0$).
Other options that do not require brother leptons are possible and can give rise to vector-like masses \cite{Craig:2016kue}.
If $\widetilde Q$ were not gauged, a brother bottom would no longer be necessary.

The equality between the couplings of $q_L$ and $\tilde t_L$ can be consistently enforced provided NLO contributions associated with extra loops of $q_L$ or gauge bosons can be neglected.
Let us study these contributions in more detail. 
Regarding the gauge couplings, the interactions of $t$ and $\tilde t$ with the EW and $\widetilde Q$ gauge bosons differ, since no exact exchange symmetry applies to the latter.
This explicit $Z_2$ breaking feeds back into the top sector at one loop, as a non-vanishing $\Delta y_L^2 \equiv y_L^2 - \tilde y_L^2$, whose size at $m_*$ can be estimated as
\beq
(\Delta y_L^2)_g = y_L^2 \frac{(A_1 g_{1}^{2} + 3 A_2 g_{2}^{2} - \tilde A \tilde{g}^{2})}{16\pi^{2}} \log\frac{\LUV}{m_{*}} \,,
\label{Z2breaktop}
\eeq
where $A_1,A_2,\tilde A$ are $O(1)$ coefficients that we cannot predict and $\LUV$ is the scale at which the differential running of $y_L$ and $\tilde y_L$ due to the gauge couplings initially arise. 
For simplicity we will take $A_1=A_2=\tilde A$ in our numerical analysis below.%
\footnote{This is a reasonable relation given the symmetries of the strong sector: they enforce $A_1=A_2=\tilde A$ exactly in the limit where the elementary gauge fields do not couple to $\mathcal{J}^{X, \widetilde X}$.}
Note that the $SU(3)_C$ gauge coupling could also be different from the $SU(3)_{\widetilde C}$ one if the colored particle content below $\LUV$ differed, for instance if the brothers of the light SM quarks were absent.
This would give rise to another term in $(\Delta y_L^2)_g$, parametrized as $A_3 (g_3^2 - \tilde g_3^2)/16\pi^2$ and generically large.
In this work we assume for simplicity that the spectrum above $m_*$ is such that $g_3(m_*) \approx \tilde g_3(m_*)$ and this contribution is absent. 

Another source of explicit $Z_2$ breaking arises from the fact that the SM bottom does not have a $Z_2$ partner (or that the presumed brother bottom does not couple to the strong sector in a $Z_2$-invariant way), as it is apparent from the embeddings in \eqs{qembed} and (\ref{qtwinembed}) and the action of the parity $\mathcal{P}$.
This gives rise to another contribution to the differential running of $y_L$ and $\tilde y_L$ that is proportional to $y_L$ itself,
\beq
(\Delta y_L^2)_y = \frac{3 B y_L^4}{16\pi^{2}} \log\frac{\LUV}{m_{*}} \,,
\label{Z2breakbottom}
\eeq
where $B = O(1)$ and in this case $\LUV$ is the scale where the differential running due to $y_L$ starts, which one might expect to be similar to that in \eq{Z2breaktop}; see Appendix~\ref{app:twin} for illustrative examples.%
\footnote{One should also presume that the boundary condition for $\Delta y_L^2$ at the scale $\LUV$ is nonvanishing, although the corresponding \emph{threshold corrections} will generically be smaller than the logarithmically enhanced contributions we have presented \cite{Weinberg:1980wa}.}

These $Z_{2}$-breaking terms add to the gauge contribution discussed in the previous section.
The implications of these effects will also be discussed in Section~\ref{sec:ewsb}.\\

There are further relevant contributions to the potential which, even if $Z_2$ symmetric, break explicitly $SO(6)$. 
The first are generated at the scale $m_*$, as in \eq{LOtoppot} but at order $O(y_L^4)$ or $O(\tilde y_L^4)$.
We compute them also via a spurion analysis, 
\begin{align} 
\label{NLOtoppot}
& \!\!\! V_{y^4}^{\textrm{UV}} = d_y y_L^4 \Big( \sum_{\psi=t,b} \Phi^T v_\psi v_\psi^\dagger \Phi \Big)^2 + d_y \tilde y_L^4 (\Phi^T v_{\tilde t} v_{\tilde t}^\dagger \Phi)^2 \\
& + c'_y y_L^4 \sum_{\psi,\psi'=t,b} |v_\psi|^2(\Phi^T v_{\psi'} v_{\psi'}^\dagger \Phi) + c'_y \tilde y_L^4 |v_{\tilde t}|^2(\Phi^T v_{\tilde t} v_{\tilde t}^\dagger \Phi)\nonumber \\
& \!\!\! = \frac{d_y}{4} \left[ y_L^4 h^4 + \tilde y_L^4 (1-h^2)^{2} \right]
+ \frac{c'_y}{2} \left[ 2 y_L^4 h^2 + \tilde y_L^4 (1-h^2) \right] \, . \nonumber
\end{align}
The term proportional to $d_{y}$ is $Z_2$ symmetric but it gives rise to both a Higgs mass and a quartic term; its NDA estimate is $6f^{4}/32\pi^{2}$. Instead the term proportional to $c'_{y}$ contains only a $Z_2$-breaking Higgs mass term; we expect it is generated at two loops, with a further suppression $(g_{*}/4\pi)^{2}$, and we neglect it in the following.

The second type of effect is IR generated, and comes from the renormalization group (RG) evolution of the potential from the scale $m_{*}$ down to the scale at which the Higgs mass is measured, that is, for instance, $m_{t}$.
The largest contribution to this running comes at leading logarithm (LL) from the light fermionic degrees of freedom coupling more strongly to the Higgs, \ie~the top and its brother.
At next-to-LL (NLL) also QCD and the brother QCD interactions should be taken into account, since they significantly contribute to the running of $y_{t}$ and $\tilde{y}_{t}$. Moreover, at NLL one also has to consider the effect of higher-dimensional operators (HDO) associated to the NGB nature of the Higgs, which modify the running of the brother top Yukawa.
Notably, the effect of RG evolution, compared to the UV threshold contributions generated at $m_*$, is logarithmically enhanced and, as already noticed in the literature \cite{Low:2015aa,Barbieri:2015aa,Greco:2016zaz,Contino:2017moj}, almost saturates the observed value of the Higgs quartic (or equivalently the Higgs mass, once the correct EW vacuum has been achieved).
Moreover, while the IR contribution is calculable solely in terms of low-energy degrees of freedom and associated observables, thus being rather model independent (up to details discussed in Appendix~\ref{app:pot}), the UV contributions depend on the spectrum of composite states. 
For these reasons, in the following we will parametrize the UV effects with free parameters, whose size we will extract (and compare with their NDA estimate) from the requirement of a phenomenologically viable Higgs potential, once the IR effect has been computed.
Examples of UV completions where the threshold contributions are calculable are two-site models \cite{Barbieri:2015aa,Low:2015aa,Contino:2017moj}, extra-dimensional constructions \cite{Geller:2014kta} and SUSY models \cite{Chang:2006ra,Craig:2013aa,Katz:2016wtw}.

Let us discuss the IR contribution in some more detail, starting with some general remarks.
There are two small parameters on which the physical Higgs mass will eventually depend: one is the leading loop expansion parameter $(g_{\text{IR}}/4\pi)^2\log(m_*^{2}/m_{\textrm{IR}}^{2})$, with $g_{\text{IR}} = \{ y_{t}, \tilde y_{t}, g_3, \tilde g_3 \}$ and $m_{\textrm{IR}} = \{ m_t, m_{\tilde t} \}$, and the other is $\xi$.
For heavy composite states with masses $m_* \sim 5$-$10 \TeV$, outside direct LHC reach, and given the constraints on $\xi$ from EW precision tests (EWPT) and Higgs couplings \cite{Contino:2017moj}, these two parameters lie in the same ballpark of 5-10\%.
It is therefore advisable to consider a joint expansion in $(\log, \xi)$, where $\log$ is a shorthand notation for the combination appearing above \cite{Contino:2017moj}.
From now on we will use the notation LL, NLL, etc.~to refer to the $\log$ expansion only, while LO, NLO etc.~will refer to the aforementioned joint expansion.
While the $\log$ expansion comes from the running of the potential, the $\xi$ expansion arises because of the non-linear properties of a NGB Higgs, encoded in the HDO. Only operators that are non-vanishing at the scale $m_{*}$ can contribute to the IR Higgs mass up to NLO. These operators can be read off our effective Lagrangian in eqs.~\eqref{kinterm} and \eqref{Yukcouplings}. In particular, on top of the HDO generated by the brother top Yukawa coupling in eq.~\eqref{Yukcouplings}, there is a single operator coming from the sigma model Lagrangian that affects the Higgs mass at NLO, that is
\beq 
\mathcal{O}_{H} = \big(\partial_{\mu} |H|^2\big)^2\,,
\label{hdo}
\eeq
with coefficient $c_H f^2/2$, where in our basis $c_{H}=1$, from \eq{kinterm}.
All other HDO that can be written with low-energy degrees of freedom, have coefficients that either vanish at the scale $m_{*}$, or cannot be predicted solely in terms of low-energy parameters. None of these operators contribute up to NLO \cite{Contino:2017moj,Greco:2016zaz}.\\

After these generic comments, let us present our results for the IR contribution to the Higgs potential, up to NLO accuracy. 
The LL term, induced by one-loop top and brother top diagrams, is given by 
\beq
V_{\textrm{LL}}^{\textrm{IR}} = \frac{3f^4}{64 \pi^2} \! \left[ y_t^4 h^4 t_{t} + \tilde y_t^4 (1-h^2)^2 t_{\tilde{t}} \right] \,, \vspace{1mm}
\label{LLpot}
\eeq
where we defined $t_{\psi}=\log(m_*^2/m_\psi^2)$ and the Yukawa couplings are evaluated at the scale $m_*$.
This potential includes the leading contribution to the Higgs quartic coupling, as well as a logarithmically enhanced Higgs mass term.
Even though the latter is negative, the corresponding minimum is at $\vev{h}^2 = \xi \approx 1/2$, which is phenomenologically excluded.
This is the reason why $Z_{2}$-breaking contributions to the Higgs mass term, already introduced above, are needed.

The potential in \eq{LLpot} gives rise to the LO contribution to the physical Higgs mass, but to NLO terms as well: from including the $O(\xi)$ correction of the Higgs kinetic term due to (\ref{hdo}), and from expressing $y_t(m_*)$ and $\tilde y_t(m_*)$ as functions of the top Yukawa coupling defined at the top mass scale $y_{t}(m_{t})$.
Besides, as it was shown in refs.~\cite{Greco:2016zaz,Contino:2017moj} (and as it happens in the SM) the LL potential leads to an overestimate of the physical Higgs mass compared to the LL-resummed one.
A better estimate (which, at least in the SM, is an underestimate of the LL-resummed result) is obtained after the RG-improvement of the Higgs effective potential at NLL.
Details on this calculation are given in Appendix~\ref{app:pot} and extensive discussions can be found in refs.~\cite{Greco:2016zaz,Contino:2017moj}.
Here we only quote the result:
\begin{widetext}
\bea
\hspace{-4mm}\displaystyle V_{\textrm{NLL}}^{\textrm{IR}} &=& \displaystyle\frac{3f^{4} h^{2}}{2048 \pi^{4}}\Bigg[ \left[-2\tilde{y}_{t}^6 \left(1-h^{2} \left(1+\frac{3 c_{H}}{2}\right)\right)+4 \tilde{y}_{t}^{4} \left(3 \left(1-h^{2}\right) y_{t}^{2}-4 \left(2-h^{2}\right) \tilde{g}_{3}^{2}\right)+2 h^{2} y_{t}^{4} \left(16 g_{3}^{2}-15 y_{t}^{2}\right)\right]t_{\tilde{t}}^{2}\vspace{2mm}\nonumber\\
\hspace{-4mm}&&\displaystyle -2 h^{2} y_{t}^{4} \left[16 g_{3}^{2}-15 y_{t}^{2}\right]t_{t} t_{\tilde{t}}+h^{2} y_{t}^{4} \left[16 g_{3}^{2}-15 y_{t}^{2}\right]t_{t}^{2} \Bigg] \,,
\label{NLLpot}
\eea
\end{widetext}
where we have retained only the terms relevant for the physical Higgs mass at NLO.
Indeed, this NLL potential contributes to the physical Higgs mass at NLO, with all its parameters evaluated at the scale $m_t$.%
\footnote{Now there is no need to run the couplings down to $m_t$, since this is a contribution of one order higher in the $\log$ expansion.
In particular, a different RG evolution of $g_3$ and $\tilde g_3$ below $m_*$, due to a different IR colored spectrum, would enter at NNLL.}
Notice that the operator $\mathcal{O}_{H}$ affects the running at NLL.
This is manifested by the appearence of the parameter $c_{H}$.
The presence of the correponding term is due to the fact that the brother top, present in the IR effective Lagrangian, has a mass term that is a relevant operator. This allows HDO like $\mathcal{O}_{H}$ to renormalize lower dimensional operators, like the brother top Yukawa coupling. This also explains why the effect enters proportional to $\tilde{y}_{t}$ (the same does not happen in the SM since the top mass term is marginal).


\section{EWSB and Higgs mass} \label{sec:ewsb}

\begin{figure*}[t]
\begin{center}
\includegraphics[width=0.443\textwidth]{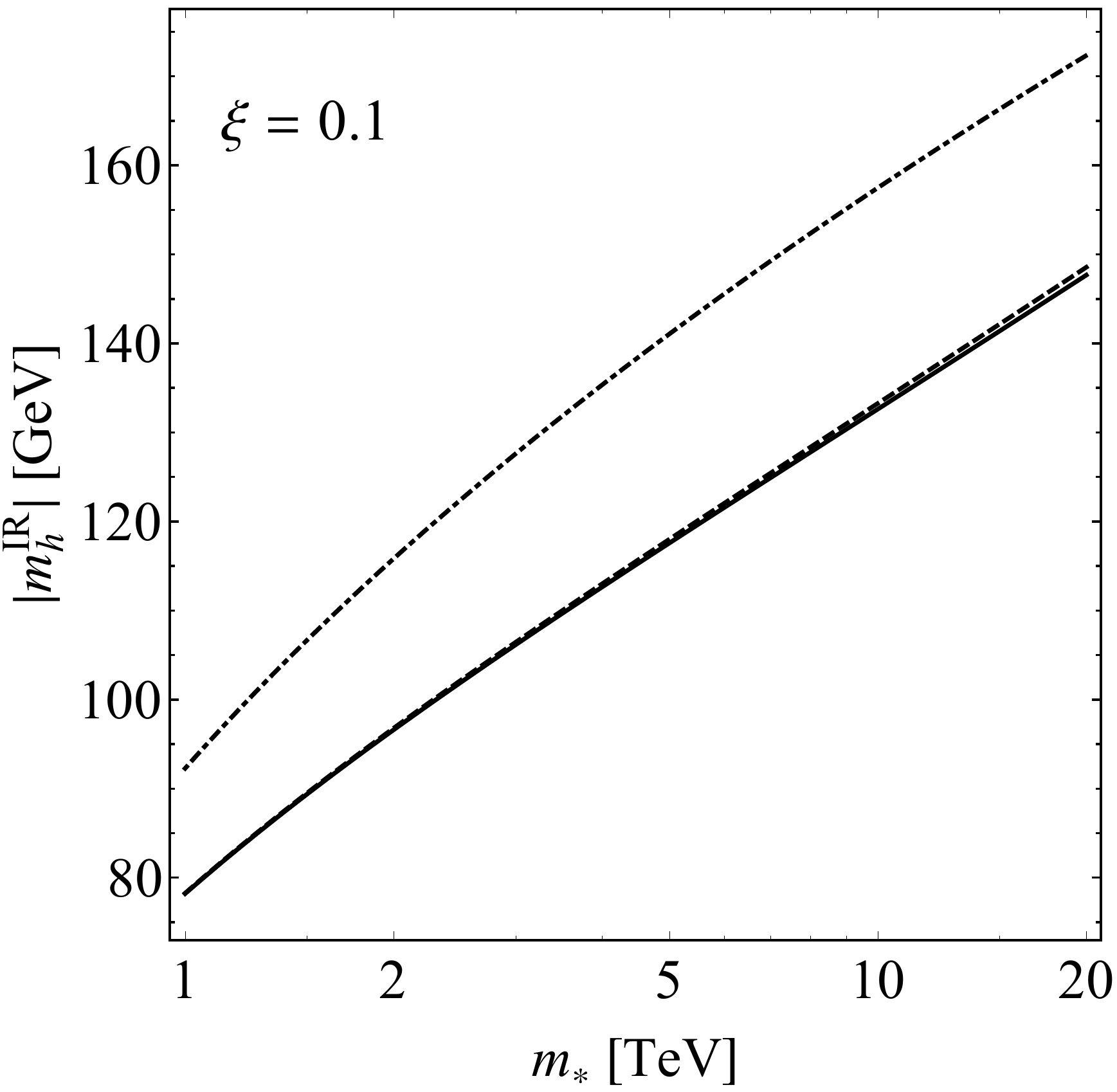}
\hspace{1.1cm}
\includegraphics[width=0.439\textwidth]{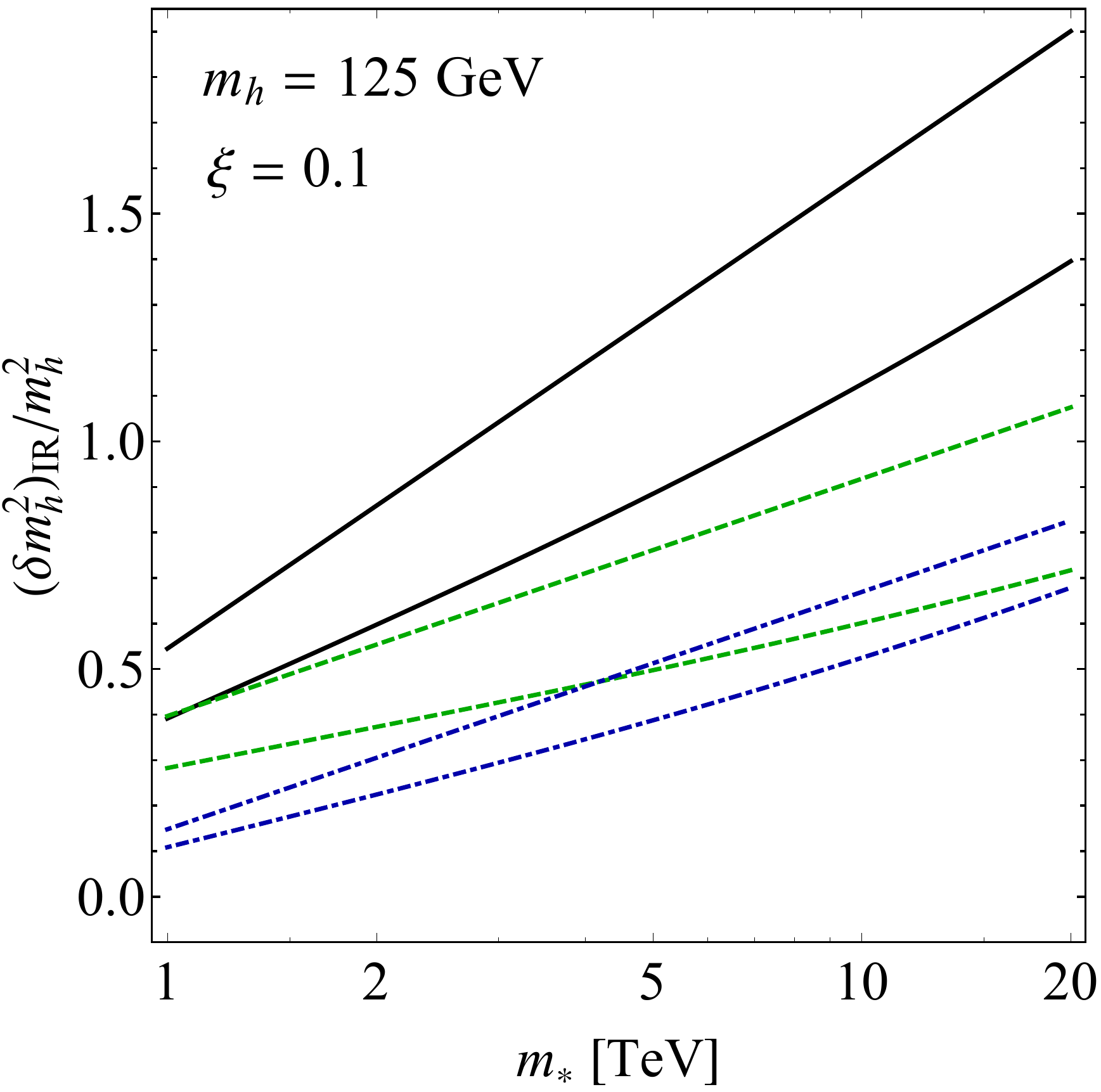}
\caption{\em Left panel: IR contribution to the Higgs mass as a function of $m_*$, at LO (dotted) and at NLO for the $SO(8)/SO(7)$ TH model (dashed) and for our brother Higgs (solid). Right panel: contributions to the Higgs mass squared normalized to its observed value: the green (dashed), blue (dot-dashed), and black (solid) lines represent the fraction of observed Higgs mass generated respectively by the top, the brother top, and both. For each color (dashing), the upper line represents the LO, while the lower line the NLO result.}
\label{fig:IRmh}
\end{center}
\end{figure*}

EWSB and the physical Higgs mass are determined by the set of contributions to the Higgs potential discussed in the previous sections: the ultraviolet terms, both the $Z_{2}$ preserving at $O(y^4)$ in \eq{NLOtoppot} and the $Z_{2}$ breaking in \eqs{gaugepot}, (\ref{Z2breaktop}) and (\ref{Z2breakbottom}), and the infrared ones, at LL in \eq{LLpot} and NLL in \eq{NLLpot}.
Since the threshold contributions are model dependent and determined by parameters we do not have full control on, we find it convenient to parametrize them as
\beq\label{UVpotparam}
V^{\textrm{UV}} = \alpha f^{4} \bar{h}^{2}+ \beta f^{4} \bar{h}^{2} (\bar{h}^{2}-1) \,,
\eeq
where $\alpha$ and $\beta$ are coefficients which we can only estimate based on NDA, and $\bar{h}^{2} \equiv h^{2}/Z_{h}\approx h^{2}(1-3y_{t}^{2}t_{t}/16\pi^{2})$ is the normalized Higgs field after taking into account the Higgs wave function renormalization.
On the contrary, the IR terms, $V^{\textrm{IR}} = V^{\textrm{IR}}_{\textrm{LL}}+V^{\textrm{IR}}_{\textrm{NLL}}$, are model independent. 
We can use this fact to extract the values of $\alpha, \beta$ required for the complete Higgs potential, $V_h = V^{\textrm{UV}} + V^{\textrm{IR}}$, to yield the correct Higgs VEV and mass. 
Trading the former for $\alpha$, the complete expression for the latter is
\begin{widetext}
\bea
\hspace{-4mm}\displaystyle m_h^2 &=& \displaystyle (m_h^{\textrm{UV}})^2 + (m_h^{\textrm{IR}})^2=16v^{2}\left[1-\xi  c_H-\frac{3 y_t^2 t_t}{16 \pi^2}\right]\beta+\frac{3v^{2}}{8\pi^{2}}\Bigg\{ \left(t_t y_t^4+t_{\tilde{t}} y_t^4\right) \left(1-\xi  c_H\right)\nonumber \\
\hspace{-4mm}&& \displaystyle +\frac{y_{t}^{4}}{32\pi^{2}}\left[ \left(\left(3 c_H-40\right) y_t^2+48 g_3^2 \right)t_{\tilde{t}}^2+ 16 \left(3 y_t^2-4 g_3^2\right)t_t  t_{\tilde{t}}+\left(3 y_t^2-16 g_3^2\right)t_t^2\right]\Bigg\}\,, 
\label{NLLmh}
\eea
\end{widetext}
where we have set $\tilde{y}_{t}(m_{*})=y_t(m_{*})$, we have expressed $y_t(m_{*})$ in terms of $y_{t} = y_t(m_{t})$ via its SM $\beta$-function, retaining only terms up to NLO order. We treated $\beta$ as a small parameter (it is loop generated) so that we have only retained terms up to order $\beta \xi$ and $\beta \log$. Equation \eqref{NLLmh} is evaluated with $y_t = y_t^{\overline{\text{MS}}}(m_{t}) \approx 0.936$ \cite{Contino:2017moj} and $c_{H}=1$. Finally, we have taken $\tilde{g}_{3}(m_{*})=g_{3}(m_{*})$ and extracted $g_{3}(m_{*})$ by running the strong coupling constant from the $Z$ mass, where $g_{3}(m_{Z}) \approx 1.22$, to $m_{*}$ according to its $\beta$-function in the SM.
The IR contribution to the physical Higgs mass, corresponding to $\beta=0$ in eq.~\eqref{NLLmh}, computed with NLO accuracy, is shown in the left panel of Figure~\ref{fig:IRmh} as a function of the cutoff of the IR effective theory, $m_{*}$, where $\xi = 0.1$ has been assumed. In the Figure we also report the NLO result in the $SO(8)/SO(7)$ TH model. The difference in the IR Higgs mass predicted in the two models is numerically irrelevant. The right panel of Figure~\ref{fig:IRmh} shows the fraction of the Higgs mass (squared) generated by the top, the brother top and their sum, at LO and NLO. Up to  the uncertainty due to the NNLO corrections (especially QCD), one can conclude that for $m_{*} \approx 5 \TeV$ the top generates almost half of $m_h^2 \approx (125 \GeV)^2$, while the brother top accounts for $\sim 40\%$. This shows, as we already anticipated, that the observed value of the Higgs mass is almost entirely saturated by the IR contribution.

\begin{figure}[t!]
\includegraphics[width=0.443\textwidth]{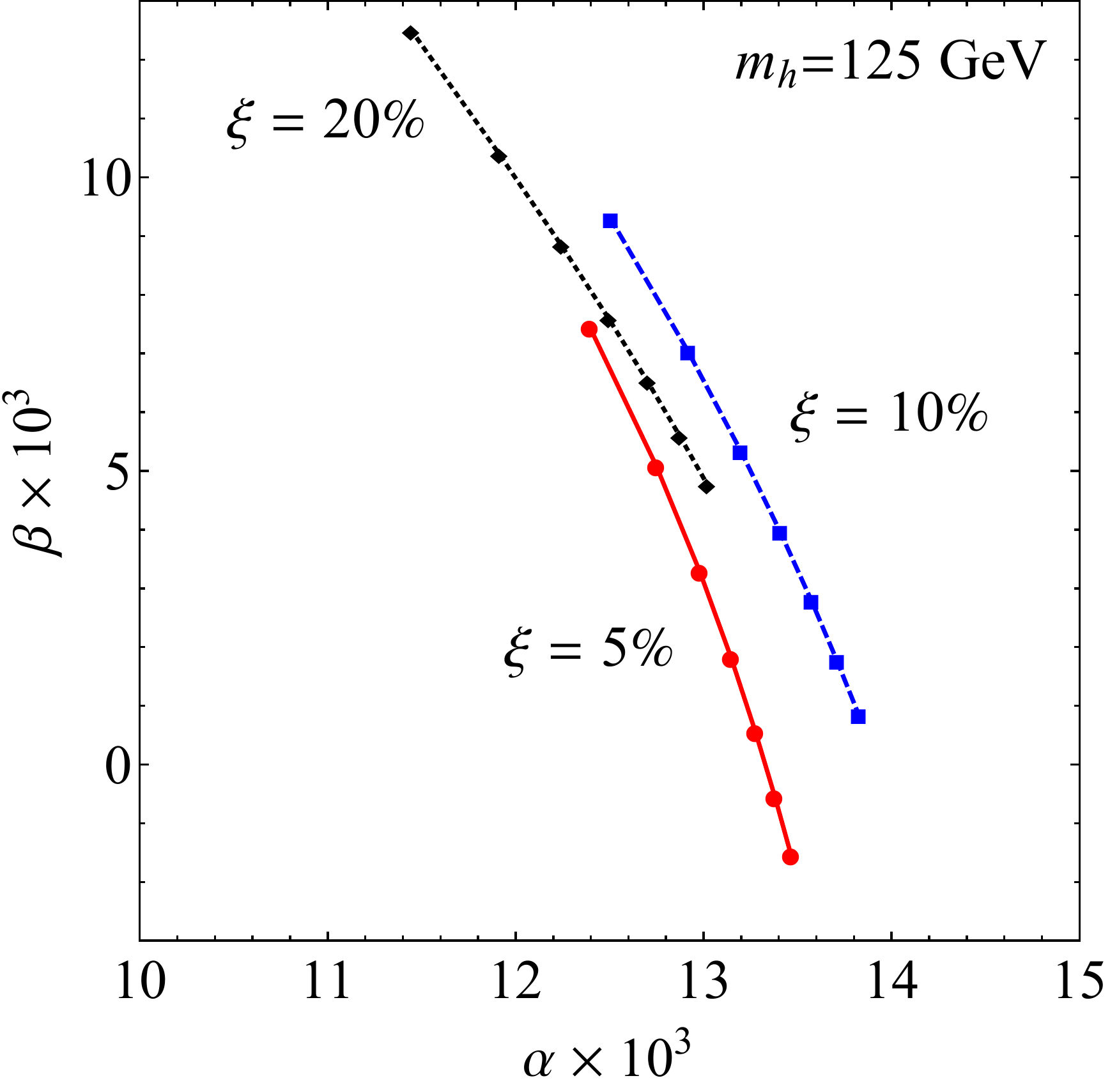}
\caption{\em Size of the threshold contributions to the Higgs potential at $m_*$, parametrized by $\alpha$ and $\beta$ in eq.~\eqref{UVpotparam}, required to reproduce the Higgs VEV (given in terms of $\xi = v^2/f^2$) and physical Higgs mass. 
For $m_{*}=g_{*}f$, the dots in the curves correspond, from top to bottom, to $g_{*}$ from $2$ to $8$.}
\label{fig:UVparam}
\end{figure}

Since this IR contribution is in general not enough to reproduce the observed Higgs mass, a non-zero UV contribution is generically required, the size of which can be read from Figure~\ref{fig:UVparam} as a value of $\beta$.
For what regards the Higgs VEV, or in other words $\vev{h}^2 = \xi$, we do not provide an analytic expression here (this can be simply obtained minimizing the full potential $V_{h}$).
Instead, we also show in Figure~\ref{fig:UVparam} the required value of the UV $Z_2$-breaking contribution $\alpha$ needed to yield a minimum of the potential at $\xi = 0.2, 0.1, 0.05$ for different values of $g_{*}=m_{*}/f$, where $m_{*}$ is the UV scale that cuts off the log's.\\

Figure~\ref{fig:UVparam} provides then simple theoretical guidance on the required size of the UV coefficients $\alpha$ and $\beta$, only as a function of $\xi$ and $g_*$.
Let us discuss the expected size of these parameters in our scenario.

From \eq{NLOtoppot} we estimate
\beq
\beta \sim \frac{3y_L^4}{32\pi^2} \approx 9 \times 10^{-3} \left( \frac{y_L}{1} \right)^2 \,,
\label{beta}
\eeq
in the right ballpark given the inherent $O(1)$ uncertainties in this estimate. 
Note that, given $y_t(m_*) \sim y_L y_R/g_*$, a light Higgs favors a fully composite right-handed top, \ie~$y_R \sim g_*$. 
Besides, one should keep in mind that the running top Yukawa coupling decreases at high energies, $y_t(5 \TeV) \approx 0.78$, hence $\beta$ in \eq{beta} could in fact be smaller.
For a discussion of the correlation between $y_L$ and EWPT see ref.~\cite{Contino:2017moj}.

Regarding the explicit $Z_2$-breaking terms, let us recall that $\alpha \neq 0$ is needed to misalign the vacuum at $\xi < 1/2$.
From \eqs{LOtoppot}, (\ref{Z2breaktop}), and (\ref{Z2breakbottom}) we have
\begin{align}
& \Delta \alpha_y \sim \frac{3 \Delta y_L^2}{32 \pi^2} \frac{m_\Psi^2 }{f^2} \\ 
& \! \approx 10^{-3} \left[ \left( \frac{g_{\textrm{eff}}}{1} \right)^2 \pm 3 \left( \frac{y_L}{1} \right)^2 \right] \left( \frac{y_L}{1} \right)^2 \left( \frac{m_\Psi/f}{4} \right)^2 \log \frac{\LUV}{m_*} \,, & \nonumber
\end{align}
where we have taken $A_1 = A_2 = \tilde A = B = 1$ and defined an effective coupling $g_{\textrm{eff}}^2 \equiv g_1^2 + 3 g_2^2 - \tilde g^2$. 
Given the theoretical uncertainty on the value of the $\widetilde Z$ gauge coupling, for our estimate we have set $g_{\textrm{eff}} = 1$. 
Comparing with Figure~\ref{fig:UVparam}, we note that $\LUV/m_* = 10^{3\textrm{-}6}$ gives a value of $\alpha$ in the right ballpark.
This estimate is, however, very crude, given the various uncertain $O(1)$ factors.
From \eq{gaugepot} we get instead
\beq
\Delta \alpha_g \sim \frac{3 g_{\textrm{eff}}^2}{128 \pi^2} \frac{m_\rho^2}{f^2} \approx 3.8 \times 10^{-2} \left( \frac{g_{\textrm{eff}}}{1} \right)^2 \left( \frac{m_\rho/f}{4} \right)^2 \!\! \,. \!\!
\label{alphag}
\eeq
Therefore the gauge contribution has a suitable magnitude too, as long as $m_\rho$ is not much larger than $f$.

Finally, note that while the size of $\Delta \alpha_y$ is determined not only by the separation between $m_\Psi$ and $f$ but also by that between $\LUV$ and $m_*$, the size of $\Delta \alpha_g$ is a more direct ``constraint'' on the ratio $m_\rho/f$, that is on the mass of the composite vector resonances, given a value of $\xi$.
If $m_\rho \gg f$, the gauge contribution to the Higgs potential would overshoot the required value of $\alpha$ to yield a minimum at $\xi$, slightly increasing the tuning beyond the minimal one, given by $2\xi$.
Nevertheless, a large ratio $m_*/f$ (and thus also $m_\rho/f$) is not expected based on perturbativity arguments. Indeed, one expects $m_\rho \lesssim m_* \sim 4 \pi f/\sqrt{N}$, for instance from $\pi \pi$ elastic scattering in $SO(N+1)/SO(N)$ \cite{Contino:2017moj}.
In summary, we find that the explicit breaking of the $Z_2$ symmetry introduced by construction in our model can potentially be of the required size.


\section{Phenomenology} \label{sec:pheno}

Given the presence of light SM-neutral states in the infrared and a relatively high cutoff scale $m_*$, our scenario gives rise to a set of phenomenological implications typical of TH or neutral-naturalness models: deviations in Higgs couplings \cite{Burdman:2014zta}, with the corresponding (IR) contribution to EWPT \cite{Barbieri:2007bh} (while UV contributions,  suppressed by $m_*$, are typically small), and possibly non-standard Higgs decays \cite{Craig:2015pha}.

However, by construction there is no massless $Z_2$ partner of the photon, nor massive $Z_2$ partners of the $W^\pm$, below $m_*$. 
Since there is no conserved twin electric charge, all our brother states can \emph{a priori} decay to SM states.
Indeed, any global symmetry associated to the brother fermions could in principle be broken, either by interactions in the elementary sector or via the strong sector. 
Several remarks in this regard are however important.
First, since the $Z_2$ partners of the $W^\pm$ are absent, there could be one conserved global symmetry for each type of brother fermion, \ie~$U(1)_{\widetilde U}$ and $U(1)_{\widetilde D}$, alluding to the number of brother up and down quarks respectively, and $U(1)_{\widetilde L}$ and $U(1)_{\widetilde N}$, for the brother leptons, if they are in the spectrum.%
\footnote{Recall that anomaly cancellation in the brother sector does not require such states.}
Second, if the strong sector preserves a given $U(1)$ brother number, its breaking by elementary interactions could be of high dimensionality and thus be suppressed by a potentially large UV scale (similar to baryon and lepton number in the SM).
Long-lived states could therefore be generic in our scenario, as in standard TH models \cite{Craig:2015pha}. 
Third, selection rules on the strong sector interactions, for instance associated with $\widetilde X$-charge if not explicitly broken, could forbid $\widetilde U$, $\widetilde D$, $\widetilde L$ or $\widetilde N$ violation.
Last but not least, if unstable, the lightest brother fermion should decay to a fermion of the SM, with possible implications for baryon and lepton number violation. 
Further details on these issues heavily depend on the spectrum of brother states.
If any of the brother fermions is a good dark matter candidate \cite{Garcia:2015loa}, included $\tilde t$, is a question that we believe deserves further study.

Finally, we can extract several phenomenological consequences from our study of the Higgs potential, that is from the Higgs mass and from the requisite of $\xi \lesssim 10 \%$ with minimal tuning.
The former indicates, according to \eq{beta}, that a significant degree of compositeness of the right-handed top is to be expected. 
Given $t_R$ couples to the strong sector as a singlet of $SO(6)$, the best way to test its compositeness is via four-top scattering \cite{Pomarol:2008bh}.
The latter instead suggests, from \eq{alphag}, that EW composite vectors are not necessarily out of LHC reach \cite{Thamm:2015zwa}.\\

We do not enter here into the issues of flavor or corrections to EWPT; these matters have been discussed in refs.~\cite{Csaki:2015gfd} and \cite{Contino:2017moj}, respectively, and we do not expect substantial changes in our scenario.


\section{Conclusions} \label{sec:conclusion}

The TH mechanism protects the pseudo-NGB Higgs potential against contributions proportional to the cutoff $m_*$, \ie~the mass of the heavy SM-charged states intrinsic of the dynamics giving rise to the Higgs.
One of the goals of this work was to explore how versatile this mechanism can really be. 
Since the largest contribution to the Higgs potential, associated with the top Yukawa coupling, is sensitive to the mass of heavy SM-colored states, for which the LHC reach is maximal, in this paper we presented an implementation of the TH protection that is only active in the top sector.

Our realization has been based on a composite Higgs model with global $SO(6)$ symmetry spontaneously broken to $SO(5)$. 
The strong sector has also been endowed with an exchange $Z_2$ symmetry such that the composite states charged under $SU(3)_C$ color behave the same as the states charged under an additional $SU(3)_{\widetilde C}$ brother color.
This paved the way to the cancellation of the leading top contribution to the Higgs potential: while the SM top couples to the strong sector via a composite operator $\Psi$, an extra elementary state, the brother top ($\tilde t_L$), couples to the strong sector via the $Z_2$-symmetric counterpart $\widetilde \Psi$.
Their contributions to the Higgs mass cancel each other out, as long as their couplings, $y_L$ and $\tilde y_L$ respectively, are the same at the relevant scale $m_*$.
The cancellation follows from the symmetry structure of $SO(6)$, which contains a discrete symmetry that controls how the couplings $y_L$ and $\tilde y_L$ break the Higgs shift-symmetries.
This is regardless of the fact that $q_L$ is a doublet while $\tilde t_L$ is not, \ie~the brother bottom is irrelevant for the TH mechanism applied only to the top sector.

This is one of the reasons we consider our model minimal. 
Another related reason is that the strong sector has a smaller symmetry breaking pattern than bona-fide TH models.
\footnote{These are based on cosets that all are $7$-spheres, $SO(8)/SO(7)$, $SU(4)/SU(3)$ or $SO(5)/SO(3)$. The only missing such type of coset, $SO(7)/G_{2}$, is being investigated by the present authors.}
In this regard our setup is similar to those presented in refs.~\cite{Poland:2008ev,Cai:2008au}, but with the important addition of custodial symmetry, crucial in order to avoid strong experimental bounds from EWPT.
The only extra NGB besides the Higgs could be eaten by gauging the associated $U(1)$, giving rise to a partial cancellation of the gauge sector contribution to the Higgs potential.
Nevertheless this is not required.

By construction our scenario breaks explicitly the $Z_2$ symmetry responsible for the TH mechanism, but in a way that still allows the cancellation of the LO Higgs potential to be at work.
We explicitly showed that NLO UV effects are under control and can give rise to the amount of $SO(6)$ and $Z_2$ breaking required for successful EWSB with minimal fine-tuning and a correct value of the physical Higgs mass.
One of the important consequences that follows from our analysis is that composite vector resonances should certainly be within reach of a future $100 \TeV$ collider, for fixed $(v/f)^2 \sim 5$-$10\%$.

Other phenomenological differences with respect to other TH models are that there is no massless twin photon and that the $Z_2$-partners of the leptons are not required by anomaly cancellation.
An additional consequence is that there are no super-selection rules associated with the conservation of twin electric charge, thus in principle all $Z_2$-partners could decay, although they could also well be (very) long-lived.

Finally, one of the interesting features of our construction is that, contrary to other $SO(N)/SO(N-1)$ models with $N\neq 6$, it can admit a fermionic UV completion \cite{Galloway:2010bp,Barnard:2013zea,Ferretti:2013kya}.
This could allow to explicitly study the dynamics giving rise to the Higgs and its interplay with the (approximate) $Z_{2}$ symmetry, in the spirit of ref.~\cite{Batra:2008jy}.


\medskip

\subsection*{Acknowledgments}

JS thanks Z.~Chacko and R.~Rattazzi for discussions. RT thanks R.~Mahbubani and R.~Rattazzi for discussion about the Higgs potential.

\smallskip

{\footnotesize
\noindent Funding and research infrastructure acknowledgments:
\begin{itemize}
\item[$\ast$] J.S. was supported in part by the ERC Advanced Grant no.267985 (DaMeSyFla) when this project was started. He would like to express a special thanks to the Mainz Institute for Theoretical Physics (MITP) for its hospitality and support.
\item[$\ast$] R.T. has been partly supported by the ERC advanced grant 669668 (NEO-NAT) and acknowledges CERN, EPFL and Corfu for hospitality during the completion of this work.
\end{itemize}
}


\appendix

\section{$SO(6)$ algebra}\label{app:so6}

The generators of $SO(6)$ in the vector representation can be written as
\beq
[ T_{KL} ]_{IJ} = - \frac{i}{\sqrt{2}} \left( \delta_{IK}\delta_{JL}-\delta_{JK}\delta_{IL} \right) \, ,
\eeq
with $I,J,K,L = 1, \dots, 6$. 
The five broken $SO(6)/SO(5)$ generators have been identified  in \eq{Phi} as $T^{\hat a} = T_{\hat a 6}$, with $\hat a = 1, \dots, 5$. 
Within the ten unbroken $SO(5)$ generators, the custodial $SU(2)_L \times SU(2)_R$ subgroup is generated by the combinations
\beq
T_{L,R}^{\alpha} = \frac{1}{\sqrt{2}} \Big( \frac{\epsilon^{\alpha \beta \gamma}}{2} T_{\beta \gamma} \pm T_{\alpha 4} \Big) \,,
\eeq
with $\alpha, \beta, \gamma = 1,2,3$.

The matrix of Goldstones is customarily given by $U(\hat \pi) = \exp \left( i \sqrt{2} \hat \pi_{\hat a} T^{\hat a} \right)$. 
In writing \eq{Phi}, we performed the field redefinitions $\pi_{\hat a} = \hat \pi_{\hat a} \sin \widehat \Pi$, with $\widehat \Pi = \sqrt{\hat \pi_{\hat b} \hat \pi_{\hat b}}$.
In this way the Goldstone matrix can be written in a compact from as $[U(\pi)]_{{\hat a}{\hat b}} = \delta_{{\hat a}{\hat b}} - \pi_{\hat a}\pi_{\hat b}/(1+\sqrt{1-\pi_{\hat a}^2})$, $[U(\pi)]_{{\hat a}6} = -[U(\pi)]_{6{\hat a}} = \pi_{\hat a}$ and $[U(\pi)]_{66} = \sqrt{1-\pi_{\hat a}^2}$.


\section{Completions} \label{app:twin}

In this appendix we discuss how the approximate $Z_2$-symmetric relation between the couplings of the left-handed top and brother top, that is $\tilde y_L \approx y_L$, can arise from more symmetric dynamics. 
We present two different proof-of-principle examples, the first based on a gauged $SO(6)$ symmetry while the second on a strong sector with a global $SO(8)$ symmetry.\\

\noindent \textbf{Example 1)} This can be regarded as the holographic version of the scenario described in the main text. 
Therefore it is based on an extra-dimensional construction, in particular a slice of AdS with UV and IR boundaries.%
\footnote{The models of refs.~\cite{Poland:2008ev,Cai:2008au} are also placed on a fifth dimension.}

The gauge symmetry in the bulk is $\mathcal{G} = SO(6) \times SU(3)_C \times SU(3)_{\widetilde C} \times U(1)_{X} \times U(1)_{\widetilde X} \times \mathcal{Z}_2$.
The unbroken subgroups of $\mathcal{G}$ at the UV and IR boundaries are $\mathcal{H}_{\mathrm{UV}} = SU(2)_L \times U(1)_{Y} \times U(1)_{\widetilde Q} \times SU(3)_C \times SU(3)_{\widetilde C}$ and $\mathcal{H}_{\mathrm{IR}} = SO(5) \times SU(3)_C \times SU(3)_{\widetilde C} \times U(1)_{X} \times U(1)_{\widetilde X} \times \mathcal{Z}_2$, with $Y = T_R^{3} + X$ and $\widetilde Q = T_\eta/\sqrt{2} + \widetilde X$. 
This pattern can be accomplished by assigning the proper boundary conditions to the 4D components of the bulk gauge fields: $A_\mu^a(+ +)$ for $T^a \in \mathcal{H} = \mathcal{H}_{\mathrm{UV}} \cap \mathcal{H}_{\mathrm{IR}}$, $A_\mu^{\bar a}(+ -)$ for $T^{\bar a} \in \mathcal{H}_{\mathrm{UV}}/\mathcal{H}$, $A_\mu^{\dot a}(- +)$ for $T^{\dot a} \in \mathcal{H}_{\mathrm{IR}}/\mathcal{H}$ and $A_\mu^{\hat a}(- -)$ for $T^{\hat a} \in \mathcal{G}/\mathcal{H}_{\mathrm{UV}} \cap \mathcal{G}/\mathcal{H}_{\mathrm{IR}}$. 
It follows that the respective 5D components have opposite boundary conditions.
In particular the $A_5^{\hat a}$ have $(+ +)$ boundary conditions, signalling the presence of scalar zero modes, the uneaten NGBs.
We define the bulk gauge couplings of $SO(6)$, $SU(3)_C \times SU(3)_{\widetilde C}$ and $U(1)_{X} \times U(1)_{\widetilde X}$ as $g_5$, $g_{5_{C}}$ and $g_{5_{X}}$, respectively. 
Then the 4D gauge couplings of the unbroken UV gauge fields are given by 
\bea
\label{gaugerel}
&\displaystyle \frac{1}{g^2} \approx \frac{L}{g_5^2} \, , \quad \frac{1}{g'^2} \approx \frac{L}{g_5^2} + \frac{L}{g_{5_{X}}^2} \, , \\ 
&\displaystyle \frac{1}{\tilde g^2} \approx \frac{L}{2g_5^2} + \frac{L}{g_{5_{X}}^2} \, , \quad \frac{1}{g_s^2} \approx \frac{1}{\tilde g_s^2} \approx \frac{L}{g_{5_{C}}^2} \,, \nonumber
\eea
where $L = R \log(R'/R)$, with $z = R$ the position of the UV boundary and $z = R'$ that of the IR boundary.
Importantly, these relations neglect UV boundary localized kinetic terms (aka threshold corrections), allowed by the UV gauge symmetry. 
Therefore, the relation between $g'$ and $\tilde g$ is an accident of the 5D construction, which holds as long as different, \ie~non-$SO(6)$ symmetric, boundary kinetic terms for $Y$ and $\widetilde Q$ are small; indeed, if the 4D gauge couplings are weak, we can expect the corrections to these relations to be small (of order one weak loop) \cite{Contino:2001si}.

Fermionic fields also propagate in the bulk, in particular the left-handed top (and bottom) $q_L$ and the brother top $\tilde t_L$, which are embedded in a bulk multiplet $\mathcal{Q}$ with the gauge quantum numbers $({\bf 6}, ({\bf 3},{\bf 1})_{2/3,0} \oplus ({\bf 1},{\bf 3})_{0,2/3})$, that is those of $\Psi_v$ and $\widetilde \Psi_v$ in Section~\ref{sec:higgs}.
Boundary conditions for the left-handed part of $\mathcal{Q}$ are chosen such that only $q_L$ and $\tilde t_L$ have zero modes, that is $(+,+)$, while the rest of the components are assigned $(-,+)$ boundary conditions. 
The right-handed components of $\mathcal{Q}$ have opposite boundary conditions.
The right-handed top $t_R$ and brother top $\tilde t_R$ are assumed to be purely IR localized, with IR quantum numbers $\mathcal{T}_R \sim ({\bf 1}, ({\bf 3},{\bf 1})_{2/3,0} \oplus ({\bf 1},{\bf 3})_{0,2/3})$.
Since $\mathcal{Q}$ decomposes on the IR brane as $\mathcal{Q}_L = \mathcal{Q}_L^{\bf 1} + \mathcal{Q}_L^{\bf 5}$ (only the left-handed components have $(+)$ IR boundary condition), a mass term is allowed on the IR brane, $m_1 \bar{\mathcal{Q}}_L^{\bf 1} \mathcal{T}_R + \mathrm{h.c.}$, giving rise to equal Yukawa couplings for the top and brother top,
\beq
y_t \approx \tilde y_t \approx \frac{g_5}{\sqrt{L}} \frac{m_1}{f}\,,
\label{yukrel}
\eeq
where we have assumed a $\mathcal{Q}$ bulk mass $m_{\mathcal{Q}} =1/(2R)$ for simplicity. 
As for the gauge fields, also for fermions the approximate relation between the couplings of the fermion zero modes holds as long no UV boundary kinetic terms for $q_L$ and $\tilde t_L$, specifically non-$SO(6)$ symmetric (and $Z_2$ breaking) ones allowed by the UV gauge symmetry, are present.
However, once again such threshold corrections are expected to be perturbatively small.

Finally, the decoupling of the unwanted UV (elementary) states by boundary conditions could be replaced by explicit 4D dynamics on the UV brane.
When trying to do such an exercise, it becomes clear that the complicated dynamics will certainly give rise to threshold effects upon integrating out the heavy scalars, fermions and vectors, which will affect the relations in \eqs{gaugerel} and (\ref{yukrel}). Nevertheless, such corrections are induced only at one-loop, so that they can be treated as a perturbation.
Besides, larger effects will be generated by RG evolution, see \eqs{Z2breaktop} and (\ref{Z2breakbottom}).
In the extra dimension, such running effects correspond to subleading corrections in the $(g_5^2/R)/(16\pi^2)$ expansion.\\

\noindent \textbf{Example 2)} We consider a composite TH model based on a global $SO(8)$ symmetry \cite{Barbieri:2015aa}.
The strong sector is also invariant under $SU(3)_C \times SU(3)_{\widetilde C} \times U(1)_{X} \times U(1)_{\widetilde X} \times \mathcal{Z}_2$ and an additional $SU(2)_{TC}$.
The subgroup $SU(2)_L \times U(1)_{Y} \times SU(3)_C \times SU(3)_{\widetilde C} \times SU(2)_{TC}$, with $Y = T_R^3 + X$, is gauged.
For the sake of simplicity, we assume that none of the twin subgroups are gauged except for twin color.
The elementary left-handed doublet $q_L$ and its twin $\tilde q_L$ are coupled to the strong sector in a $Z_2$-symmetric fashion, according to the usual embedding $q_L \in ({\bf 8},{\bf 3})_{2/3}$ of $SO(8) \times SU(3)_C \times U(1)_X$ and $\tilde q_L$ accordingly \cite{Barbieri:2015aa,Low:2015aa}.

Differently than in the standard scenario, we consider the case in which the strong sector's global $SO(8)$ is spontaneously broken to $SO(6) \times U(1)'$ by the VEV $f'$ of an adjoint $\Sigma = {\bf 28}$. 
The adjoint of $SO(8)$ decomposes as ${\bf 28} = {\bf 1}_0 + {\bf 6}_{\pm 1/\sqrt{2}} + {\bf 15}_0$ under $SO(6) \times U(1)'$, therefore the NGBs form a complex 6-plet.
The unbroken $U(1)'$ is given by $T' = (T_{\widetilde L}^3 + T_{\widetilde R}^3)/\sqrt{2}$, while the $U(1)_\eta$ within $SO(6)$ by the orthogonal combination $T_\eta = (T_{\widetilde L}^3 - T_{\widetilde R}^3)/\sqrt{2}$.

Out of the twelve NGBs, there is a complex 4-plet of $SO(4) \cong SU(2)_L \times SU(2)_R$ (a THDM charged under $U(1)'$), and two complex $SO(4)$ singlets (charged under $U(1)'$ and $U(1)_\eta$).
The former gets mass radiatively from their coupling to the SM gauge bosons.
We assume that a suitable symmetry breaking spurion lifts the latter.

The $SO(8)$ vector decomposes as ${\bf 8} = {\bf 1}_{\pm 1/\sqrt{2}} + {\bf 6}_0$ under $SO(6) \times U(1)'$. 
Upon $SO(8)/SO(6) \times U(1)'$ breaking, the SM $q_L$ remains in the vector ${\bf 6}_0$, while $\tilde q_L$ gets split: $\tilde t_L$ becomes a component of the vector ${\bf 6}_0$, while $\tilde b_L$ turns into one of charged singlets ${\bf 1}_{-1/\sqrt{2}}$.
The fact that $U(1)'$ is unbroken prevents the twins from acquiring a mass, thus $\tilde t_L$, $\tilde b_L$ and $\tilde t_R$ remain massless.
Furthermore, the breaking is such that $\tilde t_L$ remains coupled to the $SO(6) \times U(1)'$ low-energy fields with the same strength as $q_L$.
This realizes the exchange symmetry $\tilde t_{L,R} \leftrightarrow t_{L,R}$.

Without a mechanism to further break, at a lower scale $m_* \ll \LUV \sim 4 \pi f'$, $SO(6)$ to $SO(5)$, the dynamics above does not give rise to the scenario described in the main text.
To achieve this non-trivial last step, we shall imagine that the strong sector generates below $\LUV$ a composite chiral fermion $\Psi$ transforming as $(4,2)_0$ under $SO(6) \times SU(2)_{TC} \times U(1)'$, along with a set of $2n$ $SO(6) \times U(1)'$ singlets $\chi$ also in the fundamental of $SU(2)_{TC}$, \ie~$\chi \sim 2n \times (1,2)_0$.
This is then a realization of the model of ref.~\cite{Galloway:2010bp}, which breaks $SO(6)/SO(5)$ at the scale of the mass of the singlet fermions $M_\chi \chi \chi$. Therefore $m_* \sim M_\chi \ll \LUV$, represents a technically natural hierarchy.
Since in this model $U(1)_\eta$, within $SO(6)$, is not gauged, the associated NGB $\eta$ would remain massless. 
However, a mass term $M_{\Psi} \Psi^T \Psi$, allowed by the gauge symmetries, lifts it (leading to $m_\eta^2 \lesssim m_h^2/\xi$).
In this case it is natural to take $M_{\Psi} \ll M_\chi$, again because of chiral symmetry.

Finally, the Yukawa couplings for $q_L$ and $\tilde t_L$ are already generated at the scale $\LUV$, 
\beq
\sim \frac{y_L y_R}{\LUV^{d-1}} \left[ \bar Q_L (\Psi \Psi^T) t_R + \bar {\widetilde Q}_L (\Psi \Psi^T) \tilde t_R \right] + \mathrm{h.c.} \, ,
\eeq
where $d$ is the scaling dimension of $(\Psi \Psi^T)$, to be identified with $\Phi$ in \eq{Phi}. 
The Yukawa couplings of the top and the brother top are then approximately the same because of the originial $SO(8) \times \mathcal{Z}_2$ symmetry, while the IR theory reduces to the one we considered in this article.


\section{Higgs potential at NLL}\label{app:pot}

The detailed calculation of the IR Higgs potential at NLL for the composite TH based on the $SO(8)/SO(7)$ coset, using the background field method, has been presented in ref.~\cite{Greco:2016zaz}. This procedure, discussed in Section 3 of this reference, can be applied to our construction based on the $SO(6)/SO(5)$ coset, with a slight modification. While in the larger coset the running of the twin top mass is affected only by the physical Higgs, now it is also affected by the brother NGB $\eta$. In the $SO(8)/SO(7)$ TH model, the running of the twin top mass has the same form as the running of the top mass in the SM, where the contribution of the full custodial triplet of NGBs $(\pi^{0},\pi^{\pm})$ cancels out.\footnote{Notice that ref.~\cite{Greco:2016zaz} never mentions the twin NGBs, which are irrelevant for the NLL calculation, but not for the NNLL calculation, also considered in that reference.}
In our model, however, since the twin NGBs $\tilde{\pi}^{\pm}$, charged under the custodial twin $SU(2)$, are not present, the contribution of the neutral one does not cancel. This amounts to a modification of the running of the brother top Yukawa coupling. The $\beta$-function of $\tilde{y}_{t}$ in eq.~(3.11) of ref.~\cite{Greco:2016zaz} gets modified to
\beq
\beta_{\tilde{y}_{t}}\hspace{-0.5mm}=\frac{\tilde{y}_{t}(h_{c},t)}{64\pi^{2}}\hspace{-0.5mm}\left[16\tilde{g}_{3}^{2}-\tilde{y}_{t}^{2}(h_{c},t)\hspace{-0.5mm}\left(\frac{3}{Z_{\hat{h}}(h_{c},t)}\hspace{-0.5mm}-\hspace{-0.5mm}1\right)\hspace{-1mm}\frac{h_{c}^{2}}{f^{2}-h_{c}^{2}}\right],\label{backfieldbfunc}
\eeq
where $h=h_{c}+\hat{h}$ is the expansion of the Higgs field in fluctuations $\hat{h}$ around a background $h_{c}$ \cite{Greco:2016zaz}. The difference with the $SO(8)/SO(7)$ case corresponds to the $(-1)$ term in eq.~\eqref{backfieldbfunc}, that is the brother NGB contribution. Following the procedure of ref.~\cite{Greco:2016zaz} with only this modification, one gets the IR correction to the Higgs mass at NLL
\beq
\begin{array}{lll}
\displaystyle(\delta m_{h}^{2})_{\text{IR}}^{\text{NLL}}&=&\displaystyle \frac{3v^{2}t^{2}}{256\pi^{2}}\bigg[16 (g_{3}^{2}y_{t}^{4}+ \tilde{g}_{3}^{2}\tilde{y}_{t}^{4})-15y_{t}^{6}\vspace{2mm}\\
&&\displaystyle+(2+3c_{H})\tilde{y}_{t}^{6}-12y_{t}\tilde{y}_{t}^{4}\bigg]\,,
\end{array}
\eeq
where all parameters are evaluated at the scale $m_{*}$. This expression should be compared to eq.~(3.17) of ref.~\cite{Greco:2016zaz} where the NNLO $\xi t^{2}$ contribution has been neglected. We see that the only difference is in the $\tilde{y}_{t}^{6}$ contribution not proportional to $c_{H}$ (our $2$ was a $3$ in ref.~\cite{Greco:2016zaz}). This in fact turns out to be a numerically irrelevant difference, as shown in Figure \ref{fig:IRmh}. After including the LL IR correction to the physical Higgs mass up to order $\xi$, matching the expression at the scale $m_{\tilde{t}}$, including the UV contribution, and expressing all parameters as functions of the observable quantities in the IR, one gets the expression for the physical Higgs mass given in eq.~\eqref{NLLmh}.


\bibliographystyle{mine}
\bibliography{brotherhiggs}

\end{document}